\documentclass[prl,aps,twocolumn]{revtex4-2}
\usepackage[utf8]{inputenc}
\usepackage{braket}
\usepackage{amsmath}
\usepackage{amssymb}
\usepackage{mathtools}
\usepackage{pifont}
\newcommand{\cmark}{\ding{51}}
\newcommand{\xmark}{\ding{55}}
\usepackage[shortlabels]{enumitem}
\setlist{nosep}
\usepackage{amsthm}
\usepackage[hidelinks,breaklinks]{hyperref}
\newcommand{\M}{\mathcal{M}}
\newcommand{\C}{\mathcal{C}}
\newtheoremstyle{customdefi}{\topsep}{0.3\topsep}{}{}{\bfseries}{.}{.5em}{}
\theoremstyle{plain}

\theoremstyle{customdefi}
\usepackage{xcolor}

\renewcommand{\section}[1]{\par\textit{#1}.---\ignorespaces}
\renewcommand{\subsection}{\section}
\renewcommand{\subsubsection}{\section}
\usepackage{makecell}
\begin{document}
\author{Christoffer Hindlycke}
\email{christoffer.hindlycke@liu.se}
\author{Jan-Åke Larsson}
\email{jan-ake.larsson@liu.se}
\affiliation{
Department of Electrical Engineering,
Linköping University\\
581 83 Linköping, SWEDEN
}
\date{\today}
\title{Efficient Contextual Ontological Model\\ of $n$-Qubit Stabilizer Quantum Mechanics}

\begin{abstract}
The most well-known tool for studying contextuality in quantum computation is
the $n$-qubit stabilizer state tableau representation.
We provide an extension that describes not only the quantum state, but is also outcome deterministic.
The extension enables a value assignment to exponentially many Pauli observables, 
yet remains quadratic in both memory and computational complexity.
Furthermore, we show that the mechanisms employed for contextuality and measurement disturbance are wholly separate.
The model will be useful for investigating the role of contextuality in $n$-qubit quantum computation.
\end{abstract}
\maketitle

Contextuality is an important non-classical property of quantum mechanics (QM) that has been studied since the 1960s \cite{Kochen1967,Budroni2021}, while current progress in the area is connected to quantum information processing.
One tool for studying this question is the stabilizer formalism \cite{Gottesman1998a}, in particular the stabilizer state tableau representation (SSTR)  \cite{Aaronson2004} which captures the contextual behavior of the stabilizer subtheory of quantum theory.
This is widely used, both in quantum error correction and as a starting point to study properties of the quantum advantage. 
A typical question is what needs to be added to stabilizer quantum theory to achieve the quantum advantage.

However, SSTR is not an ontological model but rather a representation of the quantum states in the stabilizer subtheory, quadratic in memory and computational complexity.
An interesting question is if an ontological model, more specifically an outcome-deterministic model, can be found that is also computationally efficient.
This could then be used to study properties of the quantum advantage as compared to ontological models, rather than as compared to stabilizer QM.

The presently known outcome-deterministic models are all either non-contextual or exponential in complexity.
Perhaps the most well-known is Spekkens' toy theory (STT) \cite{Spekkens2007} from 2007, that models qubits as existing in one of four discrete ontic states, also linking predicted measurement outcomes of $Y$ to those of $X$ and $Z$. 
Though non-contextual, STT can still reproduce a number of quantum phenomena. 
This served as the stepping stone for the 8-state (cube) model \cite{Wallman2012,Blasiak2013}, wherein an additional degree of freedom is introduced for each qubit, ``decoupling'' $Y$ from $X$ and $Z$. 
Another extension is Quantum Simulation Logic (QSL) \cite{Johansson2017a,Johansson2019}, see below.
In 2019, Lillystone and Emerson \cite{Lillystone2019} proposed a contextual $\psi$-epistemic model of the stabilizer subtheory, which is outcome deterministic but exponential in memory complexity, owing to assigning an explicit phase value to each Pauli operator. 
An alternate model was also proposed which was quadratic in memory, but that model is no longer outcome deterministic.
In this article, we draw upon these previous efforts in pursuit of our goal: An  efficient, both in terms of computational and memory complexity, contextual outcome-deterministic model of the stabilizer subtheory.

We assume the reader is familiar with basics of linear algebra, the stabilizer formalism, and quantum computation \cite{Nielsen2010}.
The standard Pauli operators act on single qubits, on coordinate form
\begin{equation}
	\mathbb{I}=\begin{psmallmatrix}1&0\\0&1\end{psmallmatrix},\,
	X=\begin{psmallmatrix}0&1\\1&0\end{psmallmatrix},\, 
	Y=\begin{psmallmatrix}0&-i\\i&0\end{psmallmatrix},\,
	Z=\begin{psmallmatrix}1&0\\0&-1\end{psmallmatrix}.
\end{equation}
The \textit{$n$-qubit Pauli group} $\mathcal{P}_n$ consists of $n$-qubit Pauli operators and their respective global phase $\pm1$ or $\pm i$.
Since $iXZ=Y$, any element of $\mathcal P_n$ can be written $P=i^p \mathop\otimes_ki^{x_kz_k}X^{x_k}Z^{z_k}$ where $(x,z)$ is a \textit{binary symplectic vector}, so named because two elements $P$ and $P'$ commute iff the \textit{symplectic product}
\begin{equation}
P\cdot P'=\sum_kx_kz_k'-x'_kz_k
\end{equation}
equals 0 mod 2. The noncommutative group operation $P+P'=P''$ gives, with $x+x'=x''$ and $z+z'=z''$, 
\begin{equation}
\begin{split}
P''&
=i^{p+p'}\mathop\otimes_k i^{x_kz_k+x'_kz'_k}X^{x_k}Z^{z_k}X^{x'_k}Z^{z'_k}\\
&=i^{p+p'-P\cdot P'}\mathop\otimes_k i^{x''_kz''_k}X^{x''_k}Z^{z''_k}.
\end{split}
\label{eq:addition}
\end{equation}
This makes $\mathcal P_n$ modulo phase a symplectic vector space for which a \textit{symplectic basis} $\{M_k;C_k\}_{k=1}^n$ obeys $M_j\cdot M_k=C_j\cdot C_k=0$ mod 2 and $M_j\cdot C_k=\delta_{jk}$ mod 2.
Expansion of $M\in\mathcal P_n$ in this basis uses $m_k=M\cdot C_k$ mod 2, $c_k=M\cdot M_k$ mod 2, and binary phases $v$ and~$w$,
\begin{equation}
M=(-1)^vi^w\Big(\sum_k m_kM_k+\sum_k c_kC_k\Big).
\label{eq:basisexpansion}
\end{equation}

An $n$-qubit \textit{stabilizer state} $\ket\psi$ is uniquely determined by the subgroup $S(\ket\psi)\subset\mathcal{P}_n$ that stabilizes $\ket\psi$.
Equivalently, a stabilizer state can be obtained from $\ket0^{\otimes n}$ using only Clifford-group gates (generated by Hadamard, Phase or ``$S$'', and $CNOT$), possibly also including Pauli-group measurements.
Elements of a stabilizer subgroup are Hermitian so can be written $P=(-1)^v\mathop\otimes_ki^{x_kz_k}X^{x_k}Z^{z_k}$, and commute, so two such elements give $P\cdot P'=0$ mod 2 and 
\begin{equation}
P+P'=P''=(-1)^{v+v'-P\cdot P'/2}\mathop\otimes_k i^{x''_kz''_k}X^{x''_k}Z^{z''_k}.
\end{equation}

\section{Aim for the model}
The overall goal here is naturally to construct a model that reaches the known lower memory bound \cite{Karanjai2018}, a number of classical bits quadratic in the number of qubits, while being relatively simple to understand.
We will take inspiration from STT, and use elements of the representation of QSL.
The latter is an efficient (linear complexity, i.e., constant overhead) classical simulation framework for quantum computation, that implements one single additional resource available in quantum systems as compared to classical-bit computation, that of an additional degree of freedom of each elementary system. 
This allows for construction of quantum-like oracles, and QSL captures enough of the quantum behavior to run for example Simon's algorithm and the Deutsch-Jozsa algorithm within the oracle paradigm \cite{Johansson2019}.

QSL (and STT) achieve this by keeping track of two classical bits for each qubit in the model.
The two bits are associated with the computational degree of freedom ($z$) and the phase degree of freedom ($x$), in effect modelling a qubit using only four discrete states.
Measuring $X$ or $Z$ returns the corresponding bit, while measuring $Y$ returns the XOR of the $x$- and $z$-bit, and this makes the output deterministic given the internal state of the model.
Randomization occurs as dictated by QM: Measuring $X$ randomizes the $z$-bit to 0 or 1 uniformly, and vice versa. 
Measuring $Y$ randomizes the $x$- and $z$-bits in such a way that their XOR is unchanged ($=y$). 
Measurement outcomes are repeatable, and we obtain measurement disturbance as it occurs in QM.
Gates in QSL act on these bit values, for the Clifford group gates,
\begin{equation}
\begin{split}
H(z_h;x_h)=(x_h;z_h),\; &S(z_s;x_s)=(z_s\oplus 1;x_s\oplus z_s),\\
CNOT(z_c,z_t;\,x_c,x_t)&=(z_c,z_t\oplus z_c;\,x_c\oplus x_t,x_t)
\end{split}
\end{equation}
This makes phase kick-back manifest in the $CNOT$ gate, and many QM identities are obeyed, e.g., $HXH = Z$ and $HZH = X$. 
However, some identities fail, e.g., since the value of $y$ is given by the XOR of $x$ and $z$ in QSL we obtain $HY\!H = Y$ rather than the QM $HY\!H = -Y$.
One effect of this is that QSL (and STT) are noncontextual.
In this paper, our aim is to add contextuality.

\section{A contextual ontological model}
The main feature of QSL (and STT) is that it contains a value assignment to the symplectic basis $\{Z_k;X_k\}_{k=1}^n$, where $Z_k$ and $X_k$ are one-qubit Pauli operators acting on system~$k$. 
QSL now gives the outcome of a measurement $M$ by mod 2 summing the bit values of the symplectic basis elements contained in~$M$.

Inspired by this, the new model will still contain a value assignment to a symplectic basis for $\mathcal P_n$, but not necessarily the basis used in QSL.
We choose $\{M_k\}$ to be a basis for the stabilizer group of the quantum state of the system, so that the phase ($\pm1$) of the elements gives the predicted outcome of any Pauli measurement from that subgroup, corresponding to the value assignment.
This is not so different from SSTR, but for reasons that will become clear later, we will call this stabilizer group the \textit{measurement context} $\M$.

The second half of the symplectic basis is now needed to generate $\mathcal P_n$. In SSTR this is called \textit{destabilizer} \cite{Aaronson2004} and is used to identify measurements whose outcome should be random. 
This is where our ontological model will deviate from SSTR.
Similar to QSL we here choose $C_k$ conjugate to $M_k$, filling out the symplectic basis, under the name \textit{conjugate context} $\C$, and use the same value assignment to its elements, associating the phase to a (predicted) outcome of any Pauli measurements from that subgroup.
Measurement in the model will use three distinct steps: 
\smallskip
\begin{enumerate}[A),left=1mm]
\item \textbf{Retrieve the measurement outcome $v$.}\\
\label{meas:1}%
  Expand $M$ in the symplectic basis as in Eqn.~\eqref{eq:basisexpansion}, use $v$ as outcome, ignore $w$ because $M$ is Hermitian.
\item \textbf{Store $(-1)^vM$ as a basis element of~$\M$.}\\
\label{meas:2}%
Find $k$ so that $M\cdot M_k=c_k\neq0$ mod 2
\begin{enumerate}[i.,left=0mm]
\item If successful ($M\notin\M$), update the elements $M_j$ ($j\neq k$) for which $M\cdot M_j=c_j\neq0$ mod 2 to $M_j+M_k$, and replace $C_k$ with $M_k$.
\label{meas:2a}%
\item	Otherwise ($M\in\M$), find $k$ so that $m_k\neq0$.
\label{meas:2b}%
\end{enumerate}
Then, replace $M_k$ with $(-1)^vM$, and update the elements $C_j$ ($j\neq k$) for which $M\cdot C_j=m_j\neq 0$ mod~2 to $C_j+C_k$.
\item \textbf{Perform measurement disturbance.}\\
\label{meas:3}%
  Randomize the phase for the possibly new~$C_k$.
\end{enumerate}
\smallskip
Step \ref{meas:1} gives a well-defined deterministic map from bit values in the model to the outcome $v$. 
Step \ref{meas:2} ensures that the measurement and conjugate contexts remain a symplectic basis having updated $M_k=(-1)^vM$. 
This makes step \ref{meas:3} implement measurement disturbance with minimal complexity as only one fair coin toss is needed, mirroring measurement disturbance as it occurs in QM.

We turn now to Clifford-group gate implementation, which is straightforward: Apply the gates to all elements of the symplectic basis, including the phase according to QM identities.
Here, in contrast to QSL, the Hadamard gate acting on $Y$ will indeed result in $-Y$. 
Clifford-group gates preserve the commutation relations between Pauli operators, so the symplectic basis will remain a symplectic basis.
In coordinates \cite{Aaronson2004},
\begin{equation}
\begin{split}
&\qquad\quad H(z_h;x_h;r)=(x_h;z_h;r\oplus x_hz_h),\\
&\qquad S(z_s;x_s;r)=(z_s\oplus x_s;x_s;r\oplus x_sz_s),\\
&CNOT(z_c,z_t;\,x_c,x_t;\,r)\\
&=\big(z_c,z_t\oplus z_c;\,x_c\oplus x_t,x_t;\,r\oplus x_cz_t(x_t\oplus z_c\oplus 1)\big)\qquad\;\;\,
\end{split}
\label{eq:gates}
\end{equation}

The final part of the model is state preparation.
First choose $M_k$ so that they stabilize the initial state and mutually commute.
Second choose mutually commuting $C_k$ with random phase, that anticommute with the corresponding $M_k$ and commute with $M_j$, $j\neq k$.

The model construction obeys the \textit{Knowledge balance principle} of STT~\cite{Spekkens2007}: ``If one has maximal knowledge, then for every system, at every time, the amount of knowledge one possesses about the ontic state of the system at that time must equal the amount of knowledge one lacks.''
Step \ref{meas:3} of the measurement procedure ensures that this balance is maintained.

State preparation can also be done using Clifford group gates on $\ket{0}^{\otimes n}$, which is stabilized by $M_k=Z_k$, and one good choice of conjugate context basis with random phases $r_k$ (fair coin tosses) is $C_k=(-1)^{r_k}X_k$.
Alternatively, pick a completely random initial state and perform measurement and transformations to create the desired state.
This latter method reproduces the standard QM statement \textit{preparation is measurement}. 
(``Any measurement in quantum theory can in fact only refer either to a fixation of the initial state or to the test of such predictions, and it is first the combination of measurements of both kinds which constitutes a well-defined phenomenon'' \cite{Bohr1999}.)
Any stabilizer state can be prepared using either method.
\smallskip

\noindent\textbf{Theorem 1.}\textit{
	The model presented above is an ontological model of the $n$-qubit stabilizer subtheory.
}\smallskip

\noindent\textit{Proof.}
It suffices to show that our model gives the same predictions as SSTR \cite{Aaronson2004}.
As already observed we can use $\ket{0}^{\otimes n}$, i.e., $\{Z_k;(-1)^{r_k}X_k\}_{k=1}^n$ as the canonical initial state.
The only difference to the standard initial tableau of SSTR is that our model uses random $r_k$ whereas SSTR sets $r_k=0$ and then never uses these values.
The application of gates is identical to SSTR, see Eqn.~\eqref{eq:gates}, also implying that basis elements $C_k$ that have independent random phases before a gate array have independent random phases after the gate array.

Therefore, step \ref{meas:1} of the measurement procedure gives the same predictions as SSTR: if $M\in\M$ the outcome $v$ obtained from Eqn.~\eqref{eq:basisexpansion} equals the total \texttt{rowsum} of SSTR since both realize the group operation in $\mathcal{P}_n$, and if $M\notin\M$ the outcome $v$ will be random since it contains one or more independent fair coin tosses. 
Step \ref{meas:2} updates the basis $\{M_k;C_k\}_{k=1}^n$.
No update is done in SSTR if $M\in\M$, while our model changes basis elements but neither $\M$ nor the value assignment for $\mathcal{P}_n$, so future predictions remain unchanged. 
If $M \notin \M$ the state update of step \ref{meas:2} is identical to SSTR, with the caveat that SSTR only handles one-qubit $Z$ measurements (see the update rules for Case 1 in \cite{Aaronson2004} page 4), but this restriction can be removed.
The final step \ref{meas:3} implements measurement disturbance, which is needed in our model to maintain random independent phases for all $C_k$, so that predictions for later measurement outcomes also are exactly the same as for SSTR.
\qed

\section{Memory and computational complexity}
Storing the two contexts requires $4n^2 + 2n$ bits. 
Keeping track of interim operators and indexes during measurement updating requires at most $6n + 2n \log n + \log n +4$ bits, for a maximum concurrent memory cost of $4n^2 + 8n + 2n \log n + \log n + 4$ bits.
The model is quadratic in memory complexity, reaching the lower bound in Ref.~\cite{Karanjai2018} for classical models that simulate quantum contextuality.

Initializing the model and applying $k$ gates requires at most $4n + 2 + 16kn$ operations.
Expanding $M$ according to Eqn.~\eqref{eq:basisexpansion} requires $6n^2 + 4n$ operations. Updating the symplectic basis requires $4n^2 - n$ operations, since we may make use of many of the calculations carried out when expanding $M$. Finally, randomizing the phase of one operator requires 2 operations. 
Thus, for $k$ gates and $l$ measurements, the number of operations required is equal to $4n + 2 + 16kn + l\left(10n^2 + 3n + 2\right)$: The model is computationally efficient. Note that, for algorithms which reduce to a decision problem (where we can encode the phase value of $n-1$ qubits into ancilla qubits using consecutive $CNOT$ gates), the model is indeed quadratic in computational complexity, in the same way as SSTR.

\section{Examples of contextual behavior} From here on, we suppress the tensor notation, i.e., $X\!XY\!Z$ should be read $X \otimes X \otimes Y \otimes Z$. 
The standard example is the Peres-Mermin (PM) square  \cite{Peres1990,Mermin1990,Peres1991,Peres1993,Mermin1993,Budroni2021}. 
\begin{equation}
\begin{matrix}ZI&IZ&ZZ\\IX&XI&XX\\ZX&XZ&YY\end{matrix}
\label{eq:PM}
\end{equation}
A model that assigns noncontextual values to phases will give an even number of rows and columns that yield measurement outcomes that sum to 1 mod 2, whereas QM predicts an odd number of such rows and columns, namely the rightmost column only.
A value assignment therefore needs to be contextual (depend on measurement context, here meaning row or column), to give QM behavior.

The PM square is state-independent, but for purposes of demonstration let us here assume we begin in the state $\ket{00}$, so state preparation in our model gives the symplectic basis $\{ZI,IZ;XI,-IX\}$ (random phases 0,1 drawn by the authors).
From this starting state, let us look at measurement sequences $ZZ$;$XX$;$YY$ and $ZX$;$XZ$;$YY$, the first sequence starts with $ZZ$.
\begin{enumerate}[A),left=1mm..6mm]
\item We have $M_1+M_2=ZI+IZ=ZZ=M$ so $v=0$
\item Case \ref{meas:2b} All $c_k=0$ and $m_1=1$, so update basis to $\{ZZ,IZ;XI,-IX+XI=-XX\}$
\item Randomize the phase of $C_1$: $\{ZZ,IZ;\pm XI,-XX\}$
\end{enumerate}
Then measure $XX$.
\begin{enumerate}[A),left=1mm..6mm]
\item We have $C_2=-XX=-M$ so $v=1$
\item Case \ref{meas:2a} $c_2=1$, update to $\{ZZ,-XX;\pm XI,IZ\}$
\item Randomize the phase of $C_2$: $\{ZZ,-X\!X;\pm XI,\pm IZ\}$
\end{enumerate}
Measurement of $YY$ will find $M_1+M_2=ZZ+(-XX)=YY=M$ so $v=0$, making the outcomes from the rightmost column of Eqn.~\eqref{eq:PM} total $0\oplus1\oplus0=1$ as QM predicts.

Restarting from the initial state $\{ZI,IZ;XI,-IX\}$, the second sequence starts with $ZX$.
\begin{enumerate}[A),left=1mm..6mm]
\item We have $M_1+C_2=ZI+(-IX)=-ZX=-M$ so $v=1$
\item Case \ref{meas:2a} $c_2=1$, update to $\{ZI,-ZX;XI+IZ=XZ,IZ\}$
\item Randomize the phase of $C_2$: $\{ZI,-ZX;XZ,\pm IZ\}$
\end{enumerate}
Then measure $XZ$.
\begin{enumerate}[A),left=1mm..6mm]
\item We have $C_1=XZ=M$ so $v=0$
\item Case \ref{meas:2a} $c_1=1$, update to $\{XZ,-ZX;ZI,\pm IZ\}$
\item Randomize the phase of $C_1$: $\{XZ,-ZX;\pm ZI,\pm IZ\}$
\end{enumerate}
Here, measurement of $YY$ will find $M_1+M_2=XZ+(-ZX)=-YY=-M$ so $v=1$, making the outcomes from the bottom row of Eqn.~\eqref{eq:PM} total $1\oplus0\oplus1=0$ as QM predicts.

The measurement outcomes of $ZZ$, $XX$, $ZX$ and $XZ$ are as one would expect from the initial state. 
But importantly, the measurement outcome of $YY$ depends deterministically on what measurements are performed together with $YY$, the so-called \textit{measurement context}.
The model stores performed measurements in $\M$, hence the name.
The map to the measurement outcome of $YY$ is completely deterministic given the initial state but depends on what measurements are performed before $YY$, so the model is contextual, which is what enables it to reproduce the QM contextual behavior.
Note that while the chosen order of measurements may influence the outcomes, this influence is deterministic, and for commuting measurements the associated measurement disturbances do not change the outcomes. 

Another example is the Greenberger-Horne-Zeilinger (GHZ) paradox that uses an entangled state of three qubits with stabilizer-group generators, e.g., $-XYY$, $-YXY$, and $-YYX$; another stabilizer is $XXX=(-XYY)+(-YXY)+(-YYX)$.
These encode the correlations of the GHZ paradox, which are such that an ontological model (in the terminology used in this paper) can only reproduce these correlations if the measurement outcome at one qubit depends on what measurements are performed on the other qubits \cite{Greenberger1990}.
In this situation, such influences are usually called \textit{nonlocal}.
In our model, the GHZ state below uses three random phases $a=(-1)^r$, $b=(-1)^s$, and $c=(-1)^t$, and single system measurements give, e.g.,
\begin{align}
&\{-XYY,-Y\!XY,-YY\!X;aYII,bIYI,cIIY\}\notag\\
\xrightarrow{M=Y_1}&\{aYII,-Y\!XY,-YY\!X;\pm XYY,bIYI,cIIY\}\notag\\
\xrightarrow{M=Y_2}&\{aYII,bIYI,-YY\!X;\pm XYY,\pm Y\!XY,cIIY\}\notag\\
\xrightarrow{M=X_3}&\{aYII,bIYI,-abIIX;\pm cXYI,\pm cY\!XI,\pm IIY\}.
\end{align}
The binary outcomes sum to $r\oplus s\oplus(1\oplus r\oplus s)=1$, and so give the expected anticorrelation.
Another choice of measurement sequence gives
\begin{align}
&\{-XYY,-Y\!XY,-YY\!X;aYII,bIYI,cIIY\}\notag\\
\hspace{-3mm}\xrightarrow{M=X_1}&\{-XYY,-bcXII,IZZ;-aIXY,\pm YXY,cIIY\}\notag\\
\hspace{-3mm}\xrightarrow{M=X_2}&\{-acIXI,-bcXII,XXX;\pm XYY,\pm Y\!XY,cIIY\}\notag\\
\hspace{-3mm}\xrightarrow{M=X_3}&\{-acIXI,-bcXII,abIIX;\notag\\
&\hspace{25mm}\pm cXYI,\pm cY\!XI,\pm IIY\}.
\end{align}
The outcomes sum to $(1\oplus r\oplus t)\oplus(1\oplus s\oplus t)\oplus(r\oplus s)=0$, and so give the expected correlation.
The model is nonlocal because the measurement $X_3$ gives the outcome $1\oplus r\oplus s$ in the first case but $r\oplus s$ in the second.

\begin{figure}[tb]
	\centering
  \includegraphics{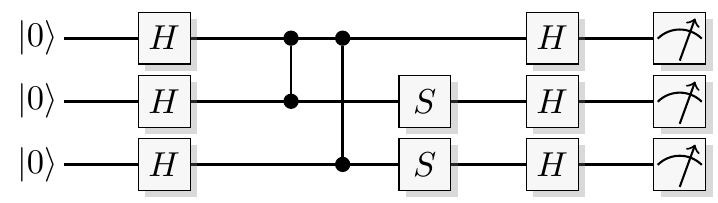}
  \caption{The quantum shallow circuits algorithm for the problem instance of Eqn.~\eqref{eq:shallowcircuit}.}
  \label{fig:shallowcircuitoracle}
\end{figure}

Our final example is the quantum shallow circuits algorithm \cite{Bravyi2018} which always succeeds when run by our model, a fact which follows immediately from Theorem 1 since the algorithm only uses (a subset of) the Clifford gates. We demonstrate the behavior for the problem instance 
\begin{equation}
	f(x) = x^TAx\text{ mod }4\text{, with }A=\begin{psmallmatrix}
  		0&1&1\\
  		1&1&0\\
  		1&0&1
  	\end{psmallmatrix}.
    \label{eq:shallowcircuit}
\end{equation}
The task is to find $z$ so that $f(x)=2z\cdot x$ mod 4 on the subset of vectors where $Ax=0$ mod 2.
The algorithm uses the circuit in Fig.~\ref{fig:shallowcircuitoracle},
and our model gives
\begin{align}
	&\{ZII,IZI,IIZ;aXII, bIXI, cIIX\}\notag\\
  \xrightarrow{HHH}&\{XII,IXI,IIX;aZII,bIZI,cIIZ\}\notag\\
  \xrightarrow{CZ_{12}}&\{XZI,ZXI,IIX;aZII,bIZI,cIIZ\}\notag\\
	\xrightarrow{CZ_{13}}&\{XZZ,ZXI,ZIX;aZII,bIZI,cIIZ\}\notag\\
	\xrightarrow{ISS}&\{XZZ,ZYI,ZIY;aZII,bIZI,cIIZ\}\\
	\xrightarrow{HHH}&\{ZXX,-XYI,-XIY;aXII,bIXI,cIIX\}\notag\\
	\xrightarrow{M=Z_1}&\{ZXX,bcZII,IYY;-aIYI,\pm XYI,cIIX\}\notag\\
  \xrightarrow{M=Z_2}&\{-abIZI,bcZII,-ZZZ;\pm IYI,\mp aXII,cIIX\}\notag\\
	\xrightarrow{M=Z_3}&\{-abIZI,bcZII,acIIZ;\pm IYI,\mp aXII,\pm IIX\}.\notag
\end{align}
Note that gates have a bounded fan-in in our model. 
The measurement output, both from our model and from QM, is with equal probability one of the solutions
\begin{equation}
z=\begin{psmallmatrix}s\oplus t\\1\oplus r\oplus s\\r\oplus t\end{psmallmatrix}
\in\left\{
\begin{psmallmatrix}1\\0\\0\end{psmallmatrix},
\begin{psmallmatrix}0\\1\\0\end{psmallmatrix},
\begin{psmallmatrix}0\\0\\1\end{psmallmatrix},
\begin{psmallmatrix}1\\1\\1\end{psmallmatrix}
\right\}.
\end{equation}

\section{Conclusion} 
We have presented an efficient contextual ontological model of stabilizer quantum mechanics.
Previously proposed models all lack at least one of efficiency, contextuality, and outcome determinism, see Table~\ref{tab:1} for a comparison. 
In addition our model is $\psi$-ontic.
Unlike Spekkens' Toy Theory \cite{Spekkens2007} and Quantum Simulation Logic~\cite{Johansson2019} our model implements contextuality for the stabilizer subtheory, and is thus able to successfully run algorithms relying on that quantum resource, such as the quantum shallow circuits algorithm as shown above.
In contrast to the models by Lillystone and Emerson \cite{Lillystone2019}, our model combines outcome determinism and efficiency.

Outcome determinism is an important difference to the Stabilizer State Tableau Representation \cite{Aaronson2004}, but note that this is more than a mere philosophical issue, as it can also be utilized in the analysis of quantum algorithms.
The Stabilizer State Tableau Representation efficiently stores the stabilizer group of a single stabilizer state and enables efficient use of Clifford-group gates and Pauli measurements, so that we can follow \textit{a single} quantum state as it is transformed, one gate after another, and subsequently measured. 
Our model in addition treats the conjugate context on almost the same footing,  storing it alongside the measurement context (that stores the stabilizer group of some selected state).
There are then several choices of stabilizer group possible in our model using elements from both contexts, so that our model enables us to simultaneously follow the behavior of all of these \textit{exponentially many} quantum states as they are transformed, one gate after another, and subsequently measured.

The model can be implemented and used in practical applications, for thousands of qubits on a modern classical computer, for example using Python~\cite{Efficientmodel2022}.
That the model can follow exponentially many quantum states using quadratic classical resources is a direct consequence of the model structure, the many possible stabilizer choices, and outcome determinism.
It is our belief that this remarkable property should prove quite helpful in enhancing our understanding of quantum algorithms.

A second property of the model is to us equally intriguing: The mechanism governing contextuality is entirely separated from that ensuring measurement disturbance. 
They are two distinct steps in the measurement update process, with no interaction between them. 
The exact ramifications of this are, at least to us, difficult to foresee; but we strongly believe this provides a very promising venue to explore further.

Finally, as our model successfully reproduces the contextual behaviour of the stabilizer subtheory while reaching the theoretical lower memory bound, it severely limits how much of the quantum advantage that can arise from stabilizer contextuality alone. 
At the very least, it suggests that to attribute the quantum advantage to contextuality one will need to delve further into the structure of contextuality itself, beyond the stabilizer subtheory. 

\begin{table}
\begin{ruledtabular}
	\begin{tabular}{cccc}
		Model & Efficient & Contextual & \makecell{Outcome-\\ deterministic}  \\[1mm]\hline
		\makecell{Stabilizer State Tableau\\ Representation \cite{Aaronson2004}} & \cmark & \cmark & \xmark \\[3mm]
		\makecell{Spekkens' Toy Theory \cite{Spekkens2007}} & \cmark & \xmark & \cmark \\[1mm]
		\makecell{Quantum Simulation \\Logic \cite{Johansson2019}} & \cmark & \xmark & \cmark \\[3mm]
		\makecell{Lillystone-Emerson \cite{Lillystone2019}} & \xmark & \cmark & \cmark \\[1mm]
		\makecell{Lillystone-Emerson\\alternate \cite{Lillystone2019}} & \cmark & \cmark & \xmark \\[3mm]
		This work & \cmark & \cmark & \cmark \\
	\end{tabular}
\end{ruledtabular}
\caption{Comparison between models. Note that Spekkens' Toy Theory is not given in efficient form in Ref.~\cite{Spekkens2007}, but can be cast in that form.}
\label{tab:1}
\end{table}

\bibliographystyle{unsrt}
\bibliography{QComputation-bibtex}

\begin{thebibliography}{10}

\bibitem{Kochen1967}
S.~Kochen and E.~P. Specker.
\newblock The problem of hidden variables in quantum mechanics.
\newblock {\em J. Math. Mech.}, 17:59--87, 1967.

\bibitem{Budroni2021}
Costantino Budroni, Ad{\'a}n Cabello, Otfried G{\"u}hne, Matthias Kleinmann,
  and Jan-{\AA}ke Larsson.
\newblock Kochen-specker {{Contextuality}}.
\newblock {\em arXiv:2102.13036 [quant-ph]}, 2021.
\newblock To appear in Rev. Mod. Phys.

\bibitem{Gottesman1998a}
Daniel Gottesman.
\newblock Theory of fault-tolerant quantum computation.
\newblock {\em Phys. Rev. A}, 57(1):127--137, 1998.

\bibitem{Aaronson2004}
Scott Aaronson and Daniel Gottesman.
\newblock Improved simulation of stabilizer circuits.
\newblock {\em Phys. Rev. A}, 70(5):052328, 2004.

\bibitem{Spekkens2007}
Robert~W. Spekkens.
\newblock Evidence for the epistemic view of quantum states: {{A}} toy theory.
\newblock {\em Phys. Rev. A}, 75(3):032110, 2007.

\bibitem{Wallman2012}
Joel~J. Wallman and Stephen~D. Bartlett.
\newblock Non-negative subtheories and quasiprobability representations of
  qubits.
\newblock {\em Phys. Rev. A}, 85(6):062121, 2012.

\bibitem{Blasiak2013}
Pawel Blasiak.
\newblock Quantum cube: {{A}} toy model of a qubit.
\newblock {\em Phys. Lett. A}, 377(12):847--850, 2013.

\bibitem{Johansson2017a}
Niklas Johansson and Jan-{\AA}ke Larsson.
\newblock Efficient classical simulation of the {{Deutsch}}\textendash{{Jozsa}}
  and {{Simon}}'s algorithms.
\newblock {\em Quantum Information Processing}, 16(9), 2017.

\bibitem{Johansson2019}
Niklas Johansson and Jan-{\AA}ke Larsson.
\newblock Quantum {{Simulation Logic}}, {{Oracles}}, and the {{Quantum
  Advantage}}.
\newblock {\em Entropy}, 21(8):800, 2019.

\bibitem{Lillystone2019}
Piers Lillystone and Joseph Emerson.
\newblock A {{Contextual}} {$\psi$}-{{Epistemic Model}} of the n-{{Qubit
  Stabilizer Formalism}}.
\newblock {\em arXiv:1904.04268 [quant-ph]}, 2019.

\bibitem{Nielsen2010}
Michael~A. Nielsen and Isaac~L. Chuang.
\newblock {\em Quantum Computation and Quantum Information}, volume 10th
  Anniversary Edition.
\newblock {Cambridge University Press}, 2010.

\bibitem{Karanjai2018}
Angela Karanjai, Joel~J. Wallman, and Stephen~D. Bartlett.
\newblock Contextuality bounds the efficiency of classical simulation of
  quantum processes.
\newblock {\em arXiv:1802.07744 [quant-ph]}, 2018.

\bibitem{Bohr1999}
N.~Bohr.
\newblock {{The}} causality problem in atomic physics (1938).
\newblock In J.~Faye and H.~J. Folse, editors, {\em The {{Philosophical
  Writings}} of {{Niels Bohr}}, {{Volume}} 4: {{Causality}} and
  {{Complementarity}}, {{Supplementary Papers}}}, pages 94--121. {Ox Bow
  Press}, {Woodbridge, CT, USA}, 1999.

\bibitem{Peres1990}
Asher Peres.
\newblock Incompatible results of quantum measurements.
\newblock {\em Phys. Lett. A}, 151(3):107--108, 1990.

\bibitem{Mermin1990}
N.~David Mermin.
\newblock Simple unified form for the major no-hidden-variables theorems.
\newblock {\em Phys. Rev. Lett.}, 65(27):3373--3376, 1990.

\bibitem{Peres1991}
A.~Peres.
\newblock Two simple proofs of the {{Kochen-Specker}} theorem.
\newblock {\em J. Phys. A: Math. Gen.}, 24(4):L175--L178, 1991.

\bibitem{Peres1993}
Asher Peres.
\newblock {\em Quantum Theory: Concepts and Methods}.
\newblock Number v. 57 in Fundamental Theories of Physics. {Kluwer},
  {Dordrecht}, 1993.

\bibitem{Mermin1993}
N.~David Mermin.
\newblock Hidden variables and the two theorems of {{John Bell}}.
\newblock {\em Rev. Mod. Phys.}, 65(3):803--815, 1993.

\bibitem{Greenberger1990}
Daniel~M. Greenberger, Michael~A. Horne, Abner Shimony, and Anton Zeilinger.
\newblock Bell's theorem without inequalities.
\newblock {\em Am. J. Phys.}, 58(12):1131--1143, 1990.

\bibitem{Bravyi2018}
Sergey Bravyi, David Gosset, and Robert K{\"o}nig.
\newblock Quantum advantage with shallow circuits.
\newblock {\em Science}, 362(6412):308--311, 2018.

\bibitem{Efficientmodel2022}
Link to~a python implementation of~our model.
\newblock
  \href{https://gitlab.liu.se/icg/efficient-contextual-ontological-model}
  {https://gitlab.liu.se/icg/efficient-contextual-ontological-model}.

\end{thebibliography}
\end{document}